\documentclass[aps,prper,longbibliography,reprint,graphicx,floatfix]{revtex4-1}

\usepackage{graphicx}
\usepackage{placeins}

\draft 

\newcommand{\etal}{et~al.}

\begin{document}

\title{Motivations of Educators for Participating in NITARP, an Authentic Astronomy Research Experience Professional Development Program}

\author{L. M. Rebull}
\email{rebull@ipac.caltech.edu}
\affiliation{Caltech-IPAC/IRSA, SSC, and NITARP, 1200 E. California Blvd., MS 100-22, Pasadena, CA 91125 USA}

\author{T. Roberts}
\affiliation{Caltech-IPAC/ICE, 1200 E. California
Blvd., MS 314-6, Pasadena, CA 91125 USA}

\author{W. Laurence}
\affiliation{Create-osity, 3187 Julies Dr., Park City, UT 84098 USA}

\author{M. T. Fitzgerald}
\affiliation{Edith Cowan Institute for Education Research, 270 Joondalup Drive, Joondalup WA 6027, Australia}

\author{D. A. French}
\affiliation{Wilkes University, 84 West South St., Wilkes-Barre, PA 18766 USA}

\author{V. Gorjian}
\affiliation{NASA/JPL, 4800 Oak Grove Dr., Pasadena, CA 91101 USA}

\author{G. K. Squires}
\affiliation{\hspace{2pt} Caltech-IPAC/ICE, 1200 E. California Blvd., MS 314-6, Pasadena, CA 91125 USA}

\date{\today}

\begin{abstract}
The NASA/IPAC Teacher Archive Research Program (NITARP) partners small groups of educators with a research astronomer for a year-long authentic research project. This program aligns well with the characteristics of high-quality professional development (PD) programs and has worked with a total of 103 educators since 2005. In this paper, surveys were explored that were obtained from 74 different educators, at up to four waypoints during the course of 13 months, incorporating data from the class of 2010 through the class of 2017. This paper investigates how participating teachers describe their motivations for participating in NITARP as evidenced in these feedback forms. Analysis of self-reported data allows a mapping onto a continuum ranging from more inward-focused to more outward-focused; there is a shift from more inward-focused responses to more outward-focused responses. This insight into teacher motivations has implications for how the educators might be supported during their year with the program.  This work provides a new way of parameterizing why educators participate in PD programs that require a considerable investment of time. NITARP, since it has many qualities of successful PD, serves as a model for similar PD programs in other STEM subjects. Likewise, the analysis method might also be useful to similarly evaluate other PD programs.  

\end{abstract}


\maketitle 

\section{Introduction}
\label{sec:intro}

Professional development (PD) that provides educators with an authentic scientific research experience can change, and more accurately frame, their impressions about the nature of scientific study, the image they hold of scientists, and effective teaching methods \cite{nstamonograph}. Authentic science, and authentic scientific research, is defined in this paper in accordance with Crawford \cite{crawford2014}, p.\ 518: ``Authentic science is a variation of inquiry teaching that aligns closely with the work of scientists, as contrasted with traditional school science laboratory exercises (commonly called ``labs'').''  Additionally, it has been demonstrated (see the NSTA position statement \cite{nstamonograph}, p.\ 2) that PD for science educators should engage them ``in transformative learning experiences that confront deeply held beliefs, knowledge, and habits of practice'' \cite{loucks2003designing,elmore2002bridging,darling1999teaching}.
  
While exact numbers are unknown, it is very likely that many teachers have not been provided opportunities to participate in authentic science experiences during their teacher preparation. For example, over half of all physics teachers do not have a major or minor in their content area (Schools and Staffing Survey 2012 as cited in Marder \cite{marder2017recruiting}). Even for teachers who majored in their science content area, not all of those programs require, or provide the option, for pre-service science teachers to participate in a deep research project while in college.  Additionally, few college science courses employ inquiry-based strategies, especially in the introductory science courses pre-service teachers take \cite{brown2006college,crawford2014,duschl2007learning,quinn2012national}.  If these courses are taught using traditional teaching strategies such as lectures, the classes do not necessarily provide opportunities for pre-service science teachers to work with authentic data.  Additionally, research shows educators teach using the pedagogical strategies experienced in the courses they took; therefore, if teachers only took lecture-based science courses, it is very likely they will also teach their classes only using lectures \cite{brown2006college,crawford2014,duschl2007learning,quinn2012national}.  Finally, teachers' perceptions and beliefs regarding the nature of science and scientific inquiry also influence their pedagogical decisions \cite{lemberger1999relationships,windschitl2004folk}.   Factors such as these lead one to expect that most teachers have not had the opportunity to engage in authentic inquiry experiences during, or prior to, their education training. We believe there exists a distinct need in the scientific education community for programs that allow teachers to experience authentic scientific inquiry.

The National Academies Framework \cite{naframework} lays out a vision where ``students, over multiple years of school, actively engage in scientific and engineering practices and apply cross-cutting concepts to deepen their understanding of the core ideas in these fields'' (p.\ 8), and that content knowledge and practices must be intertwined in learning experiences (p 11).   The Federal Science, Technology, Engineering, and Mathematics (STEM) 5-Year Strategic Plan \cite{NSTC} calls for a ``50 percent increase in the number of U.S. youth who have an effective, authentic STEM experience each year prior to completing high school'' (p. 9). Educators must engage with material to this depth before their students can.  

Quality teaching takes into account the three dimensions of learning defined in the Next Generation Science Standards (NGSS \cite{standards2013next}) as 1) science and engineering practices (SEP), 2) cross-cutting concepts, and 3) disciplinary core ideas.  As these changes are implemented, PD programs can directly provide teachers an opportunity to address the SEP steps of asking questions, defining a problem, developing and using models, planning and carrying out investigations, explaining, engaging in argument from evidence, analyzing an interpreting data, and communicating information.  PD opportunities can deepen each participant's ability to learn and model cross-cutting concepts which include finding patterns, cause and effect, scale and proportion, structure and function, and systems and models. 

As an approach to reforming STEM education, joint teacher-student research has been increasing in recent years. The teacher research experience and student inquiry-based methods have the potential to be particularly accessible in the field of astronomy. Recent changes in data access mean that astronomical data is  freely available online from many professional telescope archives, and several scientific telescope networks offer free observing time for education purposes \cite{gomez2017robotic}. However, the uptake by students and teachers in the classroom has not matched this increase in accessibility. Teachers may not access the data because they have not had the chance to learn how to use the technology or how to conduct astronomical research yet. Amongst the impediments to uptake are the inexperience of teachers with such technological and data-driven approaches, their limited knowledge about scientific inquiry, and their lack of requisite teaching skills to enact scientific inquiry in the classroom \cite{fitzgerald2015inquiry}. Because teachers are placed in the unfamiliar territory of conducting scientific research, frustrations may naturally arise. After completing an astronomy research PD, teachers reported needing more step-by-step guidance, context, and support; without such support, teachers may become frustrated and (temporarily) disengage from the process \cite{burrows2016authentic}. Teacher research experience programs can directly address all of these issues.

NITARP, the NASA/IPAC Teacher Archive Research Program (http://nitarp.ipac.caltech.edu), has provided authentic science research experiences for teachers over the last 10 years, meeting all three NGSS dimensions of learning \cite{standards2013next} and consistent with the National Academies Framework \cite{naframework}. It is a primary goal of NITARP to provide an authentic research experience which is also transformative for the participants. Participant educators are involved in a sustained year-long authentic research experience using archival data and professional astronomy tools. A group of educators is paired with a mentor astronomer, write a peer-reviewed proposal, undertake the research, write up the results and present their results at the American Astronomical Society (AAS) meeting in science poster sessions. While the program's primary audience is high school classroom teachers, it has provided this learning experience for a diverse set of educators including high school and middle school teachers, community college faculty, and informal educators.  In this paper, the words `teacher' and `educator' are used interchangeably to refer to the participants.

In addition to NITARP, well-founded teacher research programs for science educators include the Research Experience for Teachers (RET) Program supported by the National Science Foundation (NSF), STEM Teacher and Researcher program run by California Polytechnic State University (http://starteacherresearcher.org/), and the Center for Integrated Access Networks (CIAN) Research Experience for Teachers (https://blog.cian-erc.org/cian-programs-applications/research-experience-for-teachers-ret-2/). Other programs such as Math Circles \cite{kaplan1995math,donaldson2014math,white2015math,wiegers2016establishment} have expanded from students to teachers to allow educators to model mathematical thinking.  These are among a dozen or so programs that are being implemented across all sciences as the efficacy of this type of PD is realized; see Fitzgerald et al.~\cite{fitzgerald2014review} for a review of high school level astronomy student research projects, several of which focus on educators. 

The present study was undertaken specifically to address the question: How do participating teachers describe their motivations for participating in NITARP as evidenced in their feedback forms? By examining snapshots of participants' reactions to the program at four different waypoints during the NITARP year, their descriptions of their motivations for participation can be explored even though there were no questions specifically probing their motivations. One product of this research is improved teacher support during the NITARP experience. Because the demands of scientific research are high, support over frustrating junctures is important, otherwise participants may disengage from the process \cite{burrows2016authentic}. 

This study focuses in particular on the last eight years, specifically the 74 NITARP educator participants from those years. The empirical data are primarily composed of regular surveys of, and reports from, participants. This qualitative study provides important knowledge about self-reported teacher participants' needs and learning experiences in such projects and their interactions with teams, astronomy research, and their professional learning. 

Participants were given survey forms with open-ended questions at four points throughout their NITARP experience.  As a lens for analyzing and interpreting the data and to gain insight to how teachers described their motivations for participating in NITARP, a constructivist theoretical framework was used \cite{koro2009pistemological}.  Constructivism is an interpretivist theoretical framework. Within constructivism,  the researchers’ goal is to, “describe the practice” (Koro-Ljungberg et al.~\cite{koro2009pistemological}, p. 690). In this case, participants indirectly described their motivation for participating in NITARP as evidenced by their written responses to feedback forms that did not ask about motivation. Data were collected from teacher participants at four waypoints throughout NITARP. A narrative analysis was conducted on these data.  Survey responses were read and coded for evidence of whether the response indicated participants’ inward and/or outward motivation.  The content validity is given by triangulation of multiple data sources (surveys from each person at up to four waypoints), as well as member checking via participant feedback given to the researchers \cite{creswell2017qualitative}.

In this paper, an overview of NITARP is first provided (Sec.~\ref{sec:overview}), including its goals, program structure, community building, and outcomes; this program aligns strongly with literature-identified best-practices of successful PD. The demographics and data are presented in Sec.~\ref{sec:data}, including a description of the broad categories of teachers who apply to participate.  Based on the word choice the educators use in their surveys, whether they are more outward- or inward-focused in their goals and motivations for participating in the program can be identified. Sec.~\ref{sec:whyparticipate} discusses how educators are placed on that continuum, with implications for how educators can best be supported during their intensive research experience year. Sec.~\ref{sec:summary} summarizes the work presented here.

The authors of this paper include the NITARP director (LMR) and deputy (VG), mentor astronomers for NITARP teams (LMR, VG), NITARP alumni (WL, DAF), staff at IPAC involved in formal and informal education (LMR, TR, VG, GKS), education researchers (TR, WL, MTF, DAF), and professional astronomers (LMR, MTF, VG, GKS). Because the team is so heavily involved in running the program, we can use the insight provided by our experience to tell a more complete picture of the NITARP program and why teachers participate. The authors understand the context in which educators gave  responses to the feedback forms. From the perspective of running NITARP, the program is continuously adapting to the needs of teachers in general and those specifically on NITARP teams in any given year.

\section{NITARP Overview}
\label{sec:overview}

In order to answer the research question about why educators participate in NITARP, a description of what NITARP is -- its goals and how it works -- must be provided. 

NITARP's intent is to provide a long-term PD experience for teachers which enables them to experience the authentic research process. Through the program, NITARP intends to deepen educators' understanding of the nature of scientific research, and ultimately positively impact their current and future students via changes in pedagogy.

In this section, first the NITARP project is briefly described. For a more in-depth discussion of the operations model (and how it continues to evolve) via formative evaluation shaping the program to meet the teachers' needs, please see Rebull \etal~\cite{rebull2018rtsrenitarp}. Here, a typical 13-month program period is described. This section ends with a discussion of how the program aligns strongly with literature-identified best-practices of successful PD.

\subsection{High Level Summary}
\label{sec:highlevelsummary}

NITARP creates partnerships between teachers and research scientists. Small groups of educators from all over the country are paired with a professional astronomer. Teachers and mentor scientists collaborate as peers to conduct the research.  Throughout the year, teachers incorporate the experience into their classrooms and share what they have learned with other teachers, their students, and with the public.  

Participants are selected from a nation-wide application process.  There are typically at least four times as many applicants as there are positions available. Ideal applicants are already familiar with the basics of astronomy (e.g., what is a magnitude) and quantitative measures of astronomical data (e.g., what is a FITS file; see, e.g., Wells \etal~\cite{1981A&AS...44..363W}), but have no previous research experience.  Most participants teach at the high school level, but participants have also come from middle schools, community colleges and informal education settings such as planetariums and science museums. At the time of writing, the program has worked, or is working with, 103 educators from 34 states. It is estimated that these teachers reach over 22,000 students/year; this count includes students reached beyond formal high school classes, such as after-school clubs and moonlighting jobs as community college educators. 

The year-long program follows the process of research: writing a proposal, analyzing data, writing up findings, and  presenting the work at a professional society meeting. As is true of authentic research, teacher participants do not know what they will discover as their research unfolds, which surprises many of them. This insight (among others) can change teachers' perceptions of ``the scientific method'' as it is commonly taught (see, e.g., Weinburgh  \cite{weinburgh2003confronting}).  The program engages educators for at least 13 months (Jan-Jan) with many alumni choosing to remain involved over multiple years. The participants are also encouraged to involve students in the entire process.  As a result, teachers and students often learn side-by-side. Participants present their results at the January AAS conference, in the same sessions as professional astronomers, and they must `hold their own' in that domain.  They are not sequestered in a separate session where everyone knows {\em a priori} that they are high school teachers and students. 

\begin{quote}
{\em How cool for the kids to see a poster right next to theirs being presented by three university professors on one side and a graduate student on the other. } -- NITARP educator, 2010 class
\end{quote}

As of the time of writing, NITARP teams have contributed 58 science and 68 education posters to AAS January meetings (all with abstracts in ADS, the Astrophysics Data System, http://adsabs.harvard.edu/ ; entire NITARP posters are available in PDF format on the NITARP website under `publications'). NITARP teams have contributed to eight refereed papers in major astronomy journals  \cite{rebull2015infrared,rebull2013new,rebull2011new,laher2012aperture,laher2012aperture2,guieu2010spitzer,howell2008dark,howell2006first}.

\subsection{A Typical NITARP Year}
\label{sec:typicalyear}

The program format was developed and refined through the organizers' experiences over the last ten years \cite{rebull2018rtsrenitarp}. Each team consists of a mentor astronomer, a mentor teacher (a NITARP alumnus), and typically 3 new educators. The mentor educator serves as a `deputy lead', working with the professional scientist to oversee the team, and helping the new teachers with everything from logistics to science. Applications are solicited nationwide in May and due in September; strong applicants participate in a brief ($<$15 min) online interview.  Finalists are notified by early October. The teams are formed and meet for the first time in January to begin the intensive 13 months together.

To kick off the program, the program staff and the current class of educators meet for a one-day `NITARP Bootcamp,' on location at the annual January American Astronomical Society (AAS) meeting.  Half the day is spent discussing the program in general terms, and the other half working in their new teams to set goals and begin team bonding. Because this meeting is the first time that the teams have met, a good fraction of the time is spent just getting to know each other. To ensure that they have the right perspective as they launch into their projects, the timelines and expectations are reviewed at the beginning of the program. For example, feeling `stupid' is part of a scientist's job, and this is so ingrained for most scientists that it is no longer noticed. For teachers on these teams, this is usually an unfamiliar and uncomfortable feeling. During the bootcamp, one discussion explicitly focuses on how it is legitimate to feel incompetent, legitimate to not like it, and reassure participants that this is a normal part of developing as a scientist.  A 2008 article by Schwarz \cite{schwartz2008importance}, ``The Importance of Stupidity in Scientific Research'', is shared.  

\begin{quote}
{\em I also felt that [the organizers and prior class] saying it was okay to be dumb and it was okay to ask questions really helped with my comfort level.} -- NITARP educator, 2017 class
\end{quote}

Following the bootcamp, the AAS meeting begins. For most participants, this is the first time they have attended a professional scientific meeting. During at least 2 days of the AAS, they are immersed in how astronomy discourse is conducted.  They observe as astronomers share their results and interact, whether agreeing or disagreeing. And, since the Winter AAS is the biggest meeting of professional astronomers in the world (often referred to as the `Superbowl of Astronomy'), they are present as discoveries are released to the public press.

\begin{quote}
{\em This was amazing. For four straight days, if I'd thrown a rock I would have hit a scientist. Priceless. [$\ldots$] Later that day teachers who were actually teaching in my high school e-mailed me to tell me about the planets just discovered. I had to tell them I'd known about that for hours. I was there at the announcement by the scientists involved.} -- NITARP educator, 2010 class
\end{quote}

After the AAS, the educators return home and collaborate remotely to write a proposal, which is reviewed by a panel consisting of scientists and alumni educators. Teams receive comments on their proposal, and must revise it in response. The final accepted proposal is posted on the NITARP website.

Teams work collaboratively, and add students on their own timescale. Each teacher involves students on their own terms. Some educators choose to immerse themselves prior to engaging their pupils; most eagerly embrace learning side-by-side with their students. Some teachers work with large after-school or Saturday clubs; others work with smaller groups that may or may not meet during the school day.  

Through the rest of the spring, some teams hold journal clubs, focusing on papers relevant to their proposed project. Other teams begin intensive data analysis, a task made far easier in the most recent years by high-quality online collaboration tools.    

In summer, the teams each spend four days at the Californa Institute of Technology (Caltech, Pasadena, CA). The purpose of this trip is to get heavily into the data reduction and/or analysis necessary for their project. The trip includes a half-day tour of NASA's Jet Propulsion Laboratory. Each teacher may bring up to four students to this summer visit. This student limit has been empirically determined; educators bringing a larger contingent begin to operate `as a teacher' rather than `as a student.' Because NITARP's goal is educating the teachers, they need to engage in the process of research from the learner's perspective.  

\begin{quote}
{\em The best thing about the trip is the ability to get WITH all the people in one room. This is so important and just is paramount in the success of a project overall. More gets done in this week than in all the rest of the prior time leading up to the trip.} -- NITARP educator, 2012 class 
\end{quote}

\begin{quote}
{\em The structure of the program[...] sounds so simple, almost to the point of being boring: students, teachers, and an astronomer get long periods of uninterrupted time together in a room to work on a project together. But that confluence of elements is rare to the point of being unique in my science teaching experience:
(1) a clear science goal;
(2) access to an exceptional content expert;
(3) long, uninterrupted stretches of face-to-face time;
(4) necessary collaboration with strangers across a range of diverse academic backgrounds and skill sets. 
} -- NITARP educator, 2014 class
\end{quote}

During the fall, teams continue to work remotely and collaboratively to finish their project, write up results, and create two or more posters for the AAS.  One of the required two posters must be on the science results, and the other poster highlights education results;  optional additional posters expand on education results. After a year of work, the educators return to the January AAS, with students, to present their results.

Following the intensive research year, educators are asked to provide a minimum of 12 hours of related PD such as hosting workshops, seminars, labs, or giving presentations at local/regional/national teacher meetings. 

\subsection{Outcomes and Defining Success}
\label{sec:success}

The primary outcome for each team of educators is their science research results as presented in their science poster.  A secondary outcome is their education poster, which is intended to provide opportunity for reflection on the impact of the program. 

Because each team studies a new scientific question, it may seem difficult to define `success' for participants in this context. Each team uses different data, from different telescopes, sometimes at vastly different wavelengths, from archives with different interfaces, to answer different questions about objects from our Solar System to the far reaches of the Universe.  Our participants work in a variety of school environments (big/small, urban/rural, public/private), so `success' can be unique for each person even when on the same team. Sometimes, their science results are not at all what they expected, because it is authentic science. Given the data available, `success' is defined to mean that the teacher reports having gained something tangible from the experience, and the mentor astronomers can observe an increase in the participant's capabilities over the year.

\subsection{Critical Components of Successful PD}
\label{sec:successfulpd}

Quality teaching takes into account the three dimensions of learning defined in the Next Generation Science Standards (NGSS \cite{standards2013next}). One example of good astronomy PD includes helping educators directly address three-dimensional learning by integrating the Disciplinary Core Ideas of Earth and Space Science \cite{nstacoreideas} (ESS1: Earth's Place In the Universe) in conjunction with the Science and Engineering Practices and Crosscutting Concepts.  
The National Academies Framework \cite{naframework} notes that PD must change to align with the Framework's vision of coherent multi-year science education that intertwines concepts and practices.

Many PD courses are taught by facilitators who are under-trained in the content area in which they are teaching \cite{banilower2006lessons}.  NITARP is taught by content and pedagogy experts (astronomers and mentor teachers, respectively). Research shows PD lasting 50 hours or more is required for teachers to change their practice. \cite{wei2009professional}  The program runs for 13 months, from January to January; though the total amount of time varies from educator to educator, they typically devote at least 50 hours to the program within the first 2-3 months. 

Yezierski \& Herrington \cite{yezierski2013} include the following qualities of successful PD. This list below includes a description of how NITARP is strongly aligned with these qualities. 

\begin{itemize}
\item Cohort membership -- educator participants work in teams, and are welcomed into the long-term, larger community of practice \cite{lave1991situated} (CoP) consisting of alumni.
\item Collaboration with faculty -- Each team has a research mentor astronomer, and the CoP involves other mentor research astronomers.
\item Duration -- The intensive experience lasts about 13 months; alumni can be involved on longer timescales if they choose.
\item Rigor -- Participants are held to high standards; they must defend their work at the AAS meeting along with all the other professionals in the astronomy community.
\item Support -- Teams meet weekly to make progress and provide support.
\item Treated as professionals -- Teams are based on the idea that all participants, including the professional astronomer, regard one another as equal peers. Everyone works collaboratively to accomplish their science goals.
\item Accountability -- Teams meet weekly to make progress on their project, and all must defend their work at the AAS meeting.
\item Reflection -- The program promotes regular reflection on how the experience is affecting the participants through team meetings, conversations and periodic surveys.  The education poster requirement fosters and formalizes elements of their reflection. The requirement to conduct 12 hours of PD also enables reflection. 
\item Research -- Teachers and students are heavily engaged in astronomy research with archival data for a year. As part of that, the teams also become aware of current topics in astronomy research.  
\item Materials development -- Students are encouraged to participate in the entire project, to whatever degree the educator prefers (see `treated as professionals' above). Because the teacher is working next to their students, real-time materials development occurs. Rarely, however, are these materials polished and ready for posting online, although as the year progresses, individual teachers are likely to create personal lesson plans for their future classes.
\item Action research -- The primary purpose of NITARP is the astronomy research, but opportunities for action research are provided. Each team creates at least one education AAS poster, which provides opportunities for reflection on and refinement of their education practices. Some educators continue to work together as alumni to refine their teaching. 
\item Coherence -- External structure is provided for the teams each working on different projects; each team provides structure on smaller scales during the year.
\end{itemize}

Effective PD for science educators should incorporate well-founded guiding principles \cite{loucks2003designing,elmore2002bridging,darling1999teaching}, which are specifically detailed by the National Science Teachers Association (NSTA).  The NSTA Position Statement and Declaration on Professional Development in Science Education \cite{nstamonograph} states that PD programs should promote collaboration among teachers in the same school, grade, or subject. It goes on to note that educator training should be expanded so that teachers ``can benefit from national meetings and other PD opportunities that may take place away from their own school and district." \cite{nstamonograph}  In addition, ``Professional development programs should maintain a sustained focus over time, providing opportunity for continuous improvement.'' NITARP aligns well with these characteristics; it promotes collaboration, brings the teachers to national meetings and away from their own school, for a sustained amount of time.  

Recall the research question: How do participating teachers describe their motivations for participating in NITARP as evidenced in their feedback forms? Understanding the motivations of teachers has bearing on how one might support the educators through this kind of an experience. Support is explicitly listed as one of the Yezierski \& Herrington qualities of successful PD above.  Without sufficient support, teachers may disengage from the process \cite{burrows2016authentic}. 

\begin{quote}
{\em [PD] programs such as NITARP keep good teachers in the classroom teaching and leading our next generation of scientists.  Good science teachers need to be challenged, inspired, and motivated by the science they fell in love with as a student themselves.  This happens when they are able to participate and engage in current, active, real experiences such as this.  [...]  These programs make good teachers better, improve the quality of education they can deliver, and keep those highly trained, effective people in the classroom doing what they do best. } -- NITARP educator, 2016 class
\end{quote}

\section{Data}
\label{sec:data}

In this section, the data that are the focus of this analysis are first reviewed, along with the demographics of NITARP participants, followed by the specific approach used to encode the surveys.  Note that the survey questions are included in the Appendix.  Because it has bearing on interpretation of survey data, the range of teachers that apply to NITARP are discussed, specifically the four major types identified, and the ramifications for team functionality.

\subsection{Data Collection Points and Major Milestones}
\label{sec:majormilestones}

\begin{table*}
\caption{Demographics\label{tab:demographics}}
\begin{ruledtabular}
\begin{tabular}{ccccccccccp{5cm}}
Class Year & Total & \multicolumn{3}{c}{School Level} & \multicolumn{3}{c}{School Type} & \multicolumn{2}{c}{Gender} & Notes\\
 & people & HS & MS & other &  public & private & other & male & female  \\
\hline
\multicolumn{4}{l}{Spitzer years\footnote{Earliest years, funding from Spitzer education and public outreach (E/PO) mission funds. Many people repeated years during this era; the ``total unique'' row counts each person only once from 2005--2009.}}\\
2005 & 12 &10 (83\%)&2 (17\%)&0&11 (92\%)& 1 (8\%)&0&7 (58\%)&5 (42\%)& No feedback forms\\
2006 & 10 &9 (90\%)&1 (10\%)&0&9 (90\%)&1 (10\%)&0&5 (50\%)&5 (50\%) & No feedback forms\\
2007 & 16 &15 (94\%)&1 (6\%)&0&12 (75\%)&4 (25\%)&0&8 (50\%)&8 (50\%)& No feedback forms\\
2008 & 18 &15 (83\%)&3 (17\%)&0&17 (94\%)&1 (6\%)&0&10 (55\%)&8 (44\%)& No feedback forms\\
2009 & $\ldots$ &$\ldots$&$\ldots$&$\ldots$&$\ldots$&$\ldots$&$\ldots$&$\ldots$&$\ldots$& Hiatus while funding changed.\\
TOTAL UNIQUE & 34 & 30 (88\%) & 4 (12\%) & 0 & 29 (85\%) & 5 (15\%) & 0 & 18 (53\%) & 16 (47\%) & \\
\hline
\hline
\multicolumn{4}{l}{First 4 NITARP years\footnote{First years via NITARP funds. Feedback forms start being systematically collected in 2010. Many fewer people repeated years during this era, except for the mentor teachers; mentor teachers can come from any year prior, so mentor teachers generally come from the Spitzer years (and thus are counted in both the Spitzer and first NITARP years sections). Again, the ``total unique'' row counts each person only once for this section (2010--2013).}}\\
2010 & 14 & 8 (57\%) & 1 (7\%) & 5 (36\%) & 8 (53\%) & 1 (7\%) & 5 (34\%) &5 (36\%) & 9 (64\%) & \\
2011 & 11 &9 (82\%) &0&2 (18\%)&6 (55\%)&3 (27\%)&2 (18\%)&5 (45\%)&6 (55\%) &1 more dropped out\footnote{One more educator, not included in the counts here, dropped out mid-year.} \\
2012 & 19 &14 (74\%) &2 (11\%)&3 (16\%) &10 (53\%)& 6 (32\%) &3 (16\%)&7 (37\%)&12 (63\%)& \\
2013 & 18 & 14 (78\%)&0&4 (22\%)&12 (67\%)&3 (17\%)&2 (17\%)&6 (33\%)&12 (67\%)& \\
TOTAL UNIQUE &51&34 (67\%)&3 (6\%)&14 (27\%)&29 (57\%)&9 (18\%)&13 (25\%)&18 (35\%)& 33 (65\%) & \\
\hline
\multicolumn{4}{l}{Second 4 NITARP years\footnote{Second epoch of NITARP funds. Feedback forms changed to become more useful for probing impact on teachers in 2014, including adding the pre-AAS survey. Funding also decreased to half of what it was. Again, mentor teachers repeat years, and the ``total unique'' row counts each person only once for this section (2014--2017). Mentor teachers can come from any year prior, so some individuals appear once in the Spitzer epoch and once in the first NITARP epoch, or once in the first NITARP epoch and then again in the second. }}\\
2014 & 9 & 8 (89\%)&1 (11\%)&0&6 (67\%)&3 (33\%)&0&5 (56\%)&4 (44\%)& \\
2015 & 7 &5 (71\%)&2 (28\%)&0&5 (71\%)&2 (29\%)&0&4 (57\%)& 3 (43\%)& \\
2016 & 7 &6 (86\%)&1 (14\%)&0&6 (86\%)& 1 (14\%)&0&4 (57\%)&3 (14\%)& 1 more dropped out$^c$\\
2017 & 8 &7 (88\%)&1 (13\%)&0&8 (100\%)&0&0&5 (63\%)&3 (38\%)& \\
TOTAL UNIQUE  &27&22 (81\%)&5 (19\%)&0&22 (81\%) &5 (19\%)&0&16 (59\%)&11 (41\%)&  \\
\hline
\multicolumn{4}{l}{Totals for NITARP years\footnote{The NITARP years (2010-2017) are the years over which there has been a systematic collection of feedback forms, and are the focus of this work. These numbers count each person only once, whereas a mentor teacher could appear in multiple teams.}}\\
TOTAL UNIQUE & 74 & 52 (70\%) & 8 (11\%) & 10 (14\%) &48 (65\%) & 13 (18\%) & 14 (19\%)  & 32 (43\%) & 42 (57\%) \\
\end{tabular}
\end{ruledtabular}
\end{table*}

Table~\ref{tab:demographics} shows that, throughout the history of NITARP (and its immediate predecessor), 103 educators have participated. There is detailed, written survey data from 74 teachers collected over the most recent 8 years (including 2017).  Data were collected from participants at up to 4 waypoints during each NITARP year:
\begin{itemize}
\item {\bf Pre-AAS}: Before they arrive at their first AAS (initiated with the 2015 class);
\item {\bf Post-first-AAS}: Directly following the NITARP Bootcamp and their first AAS;
\item {\bf Summer}: Right after the summer work session (includes teachers and students who participate in this visit);
\item {\bf Post-second-AAS}: At the conclusion of their second AAS at which they presented their results (includes teachers and students).
\end{itemize}
In the case of mentor teachers, data has been collected during each year of their participation.  None of the educators participated on teams every year for all 8 years because their formal role as mentors is limited to 3 years maximum to allow room for others to rotate into those positions. 

Table~\ref{tab:demographics} shows that the first four years of the `NITARP years' involved nearly twice as many educators as the second four years. This reflects funding issues within the changing NASA education and public outreach (E/PO) landscape. Based on experiences with the first four years of NITARP, coupled with a better understanding of the education research literature, the surveys were substantially changed in the middle of 2014 -- after the 2014 class's first AAS but before their summer visit -- to ask different (and more specific) questions. Data from the first four years (2010-2013, but particularly 2010) are less complete than data from the most recent four years (2014-2017). The results discussed here are, by sheer numbers, weighted to the earlier years; however, by quality (and quantity) of answers per person, the results are weighted towards the later years. 

As seen in Table~\ref{tab:demographics}, over the eight NITARP years, 70\% of the participants have been the original target audience, high school classroom teachers. Moreover, 65\% of the participants are teachers in public schools. The program also has had slightly more women participate (57\%) than men (43\%). 

Because teachers have the option of bringing students along on two of the trips, there are often considerably more students than teachers on the summer and second AAS trips.  Approximately 300 students have participated in either or both of the trips with their teacher through NITARP; 240 of those students participated during the most recent eight NITARP years.

During the NITARP years, there has been 95-100\% participation from teachers on the surveys at each waypoint. Because the program's goals focus on the impact on teachers, there is an emphasis placed on obtaining surveys from every educator; these surveys are the core of the data discussed here. Students are not the focus, and student survey participation rates vary from 44-83\% per waypoint; the program rarely interacts directly with the students, leaving that role to their teacher, and efforts to obtain a high student response rate have been limited.

\subsection{Encoding}

All answers to the collected surveys were examined, and iteratively coded for emergent themes. 

For some of the themes of interest to pursue {\em a priori}, the surveys included questions that specifically probed that issue.  For instance, one theme is to determine whether participants have preconceived notions about what a scientist is, and what a scientist does (see \cite{rebull2018prperpaper2} for more on that theme).  Pointed questions were included without attempting to guide participants to an answer, but intending to capture changes in their ideas over the duration of the program.  Questions such as, ``What is real astronomy?'' or ``Did this experience change the way you thought about astronomy or astronomers?" provide illumination for their thinking and growth.  The importance of other themes (such as support during the program) emerged over several years; explicit questions to address these topics were added when the surveys were changed in mid-2014. 

However, when considering all of the survey responses in aggregate as part of this work, new recurrent themes emerged. These themes were recognized across waypoints (many surveys at the same time in the program year) and across people (up to four surveys per person per year); different surveys from the same person and different people saying the same thing at the same waypoint in the program lend validity to the results discussed here.  For those themes, there are no explicit questions and answers, but instead the themes were identified when evaluating educators' answers to the open-ended questions. Then the surveys were scrutinized again, specifically encoding for those new themes. This process prompted the realization of important themes that have guided this research and will guide future survey questions. 

For the analysis discussed in this paper, the encoding words are counted over the entirety of the survey results, once per question. For example, someone talking about students (using any words: students, children, `younger members of the team,' names of individuals) in more than one answer would receive one instance of the code word `students' for each answer given.  There are no questions that ask about students specifically; if students are mentioned by a participant twice, in two different answers, then the individual would be encoded with two instances of the word `students,' once per question. In response to a question about the best thing about the trip, if one individual wrote 500 words and another individual wrote 10 words about students, because both answers are in response to a single question, each would be encoded as a single instance of `students.'  Some additional specific examples of this are included in App.~\ref{app:encodingdetails}.

Note that NITARP teams are structured to include a mentor educator; that mentor can be on two or more teams during consecutive or non-consecutive years. The implications, then, are that a mentor teacher's answers at an AAS waypoint can apply both to the year that is finishing up (post-second-AAS) and the year that is starting (post-first-AAS). Each teacher was encoded separately, but when reporting aggregate statistics below, those AAS answers from mentor teachers continuing into the subsequent year were counted with their finishing team, not their new team; that is, a mentor teacher on 2015 and 2016 teams will appear to not have survey answers for after their first AAS in 2016, because their survey from Jan 2016 will be incorporated into the second AAS responses for their 2015 team.  It is also expected that, because of their prior NITARP experiences, the mentor teachers will come into the team already having a deeper understanding (and, thus, a smaller fractional change over the year) than their newer teammates.  Also, there is no early data to track the evolution of the mentor teachers selected out of alumni from the earliest years; they `emerge' in this analysis as already savvy in many of the themes analyzed.  For example, these teachers already experienced growth in understanding of how science works prior to the point at which they were surveyed in the context of the present work.

\subsection{Educators Right for NITARP}

Since 2005, there have been four broad `categories' of applicants, empirically noted, who wish to participate in NITARP.  A recently implemented brief online interview of the short-list of applicants provides insight before making offers to the finalists. With the limited resources available to NITARP, educators must be selected so as to provide the largest `lever arm' per dollar spent or hour invested in their training -- that is, gain the most from the experience and share the experience widely.  Below, four broad applicant types are identified:
\begin{enumerate}
\item {\bf Ideal candidate.} In most years, the program has had more ideally qualified educator applicants than there are funded positions available. These educators are ready to do research but have not yet done it. They are already using data with students, preferably also inquiry (or modeling) labs and techniques, and are skilled with computers. They have a working knowledge of college level astronomy. Operationally, teachers need to be able to handle a high rate of email, as well as a fluctuating time commitment, over 13+ months, without monetary compensation. They need not currently be teaching astronomy or a student research class; participants have also been math, chemistry, or Earth science teachers. However, if they are not teaching astronomy, then they need to have enough flexibility to incorporate astronomy into their classroom, and/or run astronomy programs outside of school hours. They must somehow share their experiences with their community (students, teachers, amateur astronomy clubs, district, region, etc.). 
\item {\bf Overqualified.} Educators that either already have a M.S.\ or Ph.D.\ in the physical sciences (or already have done research, attended AAS meetings, and presented posters or written papers) are overqualified for this teacher research experience. These educators probably already understand how scientific research works. While these teachers are likely to enjoy NITARP, the fractional gain that the program could give them is low compared to the ideal candidate who has not yet explored how research works.  Some apply for the program citing the hope that NITARP can help them better integrate data into their classroom; these applicants are referred to NSTA and AAPT resources.
\item {\bf Underqualified.} If applicants do not have a working knowledge of college level astronomy prior to the program, they are less prepared to jump directly into research; too much time could be required to teach them basic vocabulary and background information. NITARP has limited resources and so cannot train everyone from the ground up; teachers who are not yet fluent in college-level astronomy need to become fluent via other opportunities.  In some cases, educators feel that they are ready, but in talking to them, it becomes apparent that they have not yet mastered the basics.  In general, new teachers are also not ideal candidates because freshly trained teachers need to gather a great deal of experience about classroom management and administrative policies; they need to master those before they can easily incorporate information from NITARP into their curriculum. 
\item {\bf ``Experience collectors.''} This type of applicant appears to love to add to their resume after they have completed a NASA program, but are less enthused about actually doing the project. This candidate may initially present as perfectly qualified, but when it is time to really work, it turns out they don't have the necessary drive.  The finalist interview process recently implemented has helped eliminate this type of candidate. 
\end{enumerate}

In practice, during the year, it can be hard to separate experience collectors from people who really are ideal candidates but are paralyzed because they are confronted with so much new material that they do not (yet) know how to tackle it. 

\label{sec:fish}
Some educators experience learning roadblocks when they realize they have, perhaps for the first time, encountered a learning situation that is beyond what they know and/or how they know to engage in the process of learning --  what could be characterized as a `big fish/small pond' phenomenon. For some who are used to being the only leader or star, there is an adjustment process as they join a team filled with highly successful educators and accomplished scientists. When confronted with this hurdle, a few NITARP participants have shut down and effectively removed themselves from participation; some never fully realize they have to work harder than ever before just to keep up. 

As a result of noticing these patterns, the Bootcamp has come to include a forthright discussion of how overwhelming this program can be, and suggestions on how they may overcome these hurdles. For example, the likelihood of feeling overwhelmed often during the year is discussed, and program personnel point out multiple times during the subsequent year that this can happen and how they can become active members of a high-functioning team even when they feel overwhelmed. Explicitly pointing out potential pitfalls seems to have ameliorated issues of broken teams and educators `giving up' entirely.

\begin{quote}
{\em Usually things are pretty easy for me, but not this. } -- NITARP educator, 2016 class.
\end{quote}

\begin{quote}
{\em The social support of my group is also helpful when I'm feeling completely lost; we don't get involved in our egos and [do] care about helping each other. } -- NITARP educator, 2016 class.
\end{quote}


\section{Findings: The Focus of Educator Participation}
\label{sec:whyparticipate}

Our primary research question is: How do participating teachers describe their motivations for participating in NITARP as evidenced in their feedback forms? In this section, we explore why these educators came to NITARP, and why they chose to devote so much unpaid time. In doing so, insights into how and when participants might need additional support, as well as a better understanding will be achieved regarding the `right' candidates to select in the future. 

There are myriad reasons why a teacher may choose to participate in PD that requires a considerable commitment of time and resources over a sustained period. Examining these motivations has helped to understand educators' stance, or approach to their learning, which in turn impacts their participation in the program. These patterns are discussed in this section because understanding their stance leads to more cognizance of the changes they experience during the program (also see Rebull \etal~\cite{rebull2018prperpaper2}). However, it is important to note that these findings were emergent and the result of serendipitous data that permitted investigation of emergent patterns. As stated earlier in this paper, scientists often do not know what they will discover as their research unfolds, and this has certainly been true here. 

In this section, encoding of words used by all the NITARP educators are examined in the entirety of their (encoded) survey results. From the distribution of word selection in the encoded surveys, it was found that the educators can be mapped onto a `continuum' or range of focus.  PD is undertaken to meet the needs of a teacher, at varying stages of development in their career and personal growth. There is no judgment assigned to location on the continuum, but it reveals a critical point -- that educators may be participating to meet a personal need which they recognize and are able to verbalize, but over the course of the program, they often discover additional benefits that fill gaps which were previously unknown to themselves.

\subsection{Inward (Personal Learning Goals) and Outward (Teaching Goals)}
\label{sec:inandout}

One end of the continuum is more `inwardly focused'; these participants express a strong desire for personal learning -- new science, new skills, collaborating with like-minded colleagues, gaining access to opportunities (within and beyond NITARP), etc. These may be people who feel isolated in their home schools, either because the school itself is small, or there are few science teachers, or perhaps they see few others at their school as interested in the same material. This end of the continuum includes those who are searching for the intrinsic reward of learning which comes with tackling increasingly challenging projects.   

\begin{quote}
{\em Any time I can meet with other educators who teach what I teach, I benefit.  Having the opportunity to get new ideas from my peers and discuss projects, activities, and strategies helps me to grow and keeps me from becoming stagnant in my teaching. } -- NITARP educator, 2011 class
\end{quote}

\begin{quote}
{\em ...it inspired me. My colleagues and I are teachers. Just as you, scientists, are developing your portfolio and skills as teachers, we need to do a better job of also being scientists. We all need to attempt the Feynman professional duality. You have reached out to embrace teaching and given us the means to reach out and embrace science.} -- NITARP educator, 2016 class
\end{quote}

\begin{quote}
{\em For the teacher, connecting with working scientists and networking with other colleagues has immeasurable value.  I plan to utilize these relationships and potentially other projects spawned by them for years to come.} -- NITARP educator, 2012 class
\end{quote}

\begin{quote}
{\em Being with the people drawn together at this type of meeting helps me as a teacher to see what is needed from me in prepare and present to my students as the current world of science.  I also get to see and experience things that make my own brain start clicking and re-engage that wonder and questioning part of me that made me love science and want to go into science as a kid.  I came away with many new ideas, new contacts to offer me support in my teaching and research, and a renewed enthusiasm for improving my teaching and my own understanding of astronomy.} -- NITARP educator, 2013 class
\end{quote}

\begin{quote}
{\em The best thing about the trip was the chance to interact with others who are trying to do the same things that I am trying to do.  No one else around me tries to do student research (even though I have tried to get other teachers involved), not in my district nor in any of the surrounding ones.   It was great to spend time with other teachers (and their students) who are trying to accomplish the same things that I am trying to do.} -- NITARP educator, 2013 class
\end{quote}

\begin{quote}
{\em The BEST subject-area professional development experience I've had in 25 years BY FAR, and one of the most intellectually stimulating experiences I've had in years. I lie awake at night thinking about data. } -- NITARP educator, 2014 class 
\end{quote}

The other end of the continuum is more `outwardly focused' and expresses interests related to teaching goals or student benefits, e.g., student gain -- helping students get involved in authentic research and/or International Science and Engineering Fair (ISEF) projects, watching students gain confidence, etc. To a lesser extent, some of the outward-focused teachers mention how excited they are to share their experience with other educators. These educators are not participating primarily for themselves; they participate because they want to see their students grow and change, or see changes in other educators' practices. Often they list getting students involved in authentic research as a primary goal.

\begin{quote}
{\em NITARP will expand my ability to offer exciting and meaningful educational opportunities to students in my classes and to interested and able students in my own school, and such students in many secondary schools in my geographic area.} -- NITARP educator, 2017 class
\end{quote}

\begin{quote}
{\em [The best thing was] Watching the students interact with each other and with the science. It was great to see kids from different schools working with each other. It was also really great to see the students embracing the science. They asked our astronomers thoughtful questions that showed they were thinking about the process and the science. } -- NITARP educator, 2012 class
\end{quote}

\begin{quote}
{\em I cannot say enough positives about the NITARP experience for the participating students. They have had the opportunity to learn and grow and see science applied in authentic research projects while working with some of the coolest scientists around! It has allowed me to grow as a teacher and researcher and be able to share my insight and newfound knowledge with students and peers. } -- NITARP educator, 2010 class
\end{quote}

\begin{quote}
{\em It was most rewarding to watch my students gain confidence in science and to shed some self-doubt.} -- NITARP educator, 2013 class
\end{quote}

\begin{quote}
{\em It was a delight to watch the students explain the poster -– usually followed by shock as the person listening noticed they were middle and high school students! Here is to the next generation -– they are amazing. } -- NITARP educator, 2013 class
\end{quote}

\begin{quote}
{\em The best thing about the trip for me was watching how much my students' learning had evolved over time.  So much of what we have done so far had a really steep learning curve, and it was really great to see them communicate the details of our project at the end of the Caltech visit. } -- NITARP educator, 2012 class
\end{quote}

\begin{quote}
{\em Best thing:  Seeing kids WORK!  And getting confused.  I liked working together with students to accomplish the tasks.} -- NITARP educator, 2014 class
\end{quote}

In short, `inwardly focused' goals are more intrinsic, or self-related to their own learning or personal gain, whereas `outwardly focused' goals are 1people-related' and directed toward helping others.  Both goals might outwardly look the same but originate from different intent.   Both ends of this continuum are worthy justifications for participating, and participants from anywhere within the continuum are successful (where success is defined above). In practice, participants fall over the whole range, and moreover, they move along the continuum during the year. 

\subsection{Placement on the Continuum}
\label{sec:placementonthecontinuum}

In order to assess the range of the focus continuum, and to see how people move during the year, survey answers from each educator's responses at each of the four waypoints were encoded with any of the following five words; included are a list of example statements (provided in parentheses) that are the simplified essence of the sentiment, not direct quotes: 
\begin{itemize}
\item {\em Students} to indicate emphasis on student benefits. (I want to help students conduct real research; I want to get more authentic research into my classroom; the most important thing was watching my students present their findings to the team during the summer visit; the best thing was watching my students gain in confidence as they presented our poster.)
\item {\em Teachers} to indicate that they are talking about sharing with other teachers, not those in NITARP. (I can't wait to share this with my fellow teachers when I get home; I am already thinking about how to share this with other teachers in my district.)
\item {\em Team} to indicate emphasis on the team effort, working together on a project with a clear purpose, or reporting issues with a team. (I can't wait to meet my team; I can't believe how fast we bonded; the visit really made us gel as a team; it's hard to work in a team if someone isn't pulling their weight; I will miss my team.)
\item {\em Colleagues} to indicate someone who describes the importance of and/or how they personally benefit from finding and/or working with and/or being inspired by like-minded colleagues from across the country. Such colleagues can be found within NITARP or just as part of the AAS experience.  (I met so many people doing what I'm trying to do and I can learn from their experiences; I really enjoy meeting other teachers like me because I learn from them; I really enjoy working with other like-minded people because it helps me grow.) 
\item {\em Self} to indicate they are focused on their own experience and/or personal impacts of this type of PD. (I liked going to these talks; I enjoyed the tour of JPL; I really improved my ability to use Excel.) 
\end{itemize}

We emphasize the point again that this represents an encoding of the words used by educators in their responses into these five words, not the usage of these five words by the educators per se.  (Recall specific examples of the encoding are included in App.~\ref{app:encodingdetails}.)

Within the inward/outward focused paradigm described above, {\em students} and {\em teachers} are outward-focused, and {\em colleagues} and {\em self} are inward-focused.  {\em Team} is both and neither inward and/or outward focused. Given the way that it is encoded for this work, it includes sharing within the team (with both educators and students), but does not encompass sharing with other students or teachers external to NITARP. It is not focused on one's personal gains, but it is of direct benefit to each person on the team, including the teacher who is writing the comments encoded. It speaks to bonding and progress on the project, and how `present' in their minds the whole aspect of the team effort is. 

\begin{figure}
\includegraphics[width=3in]{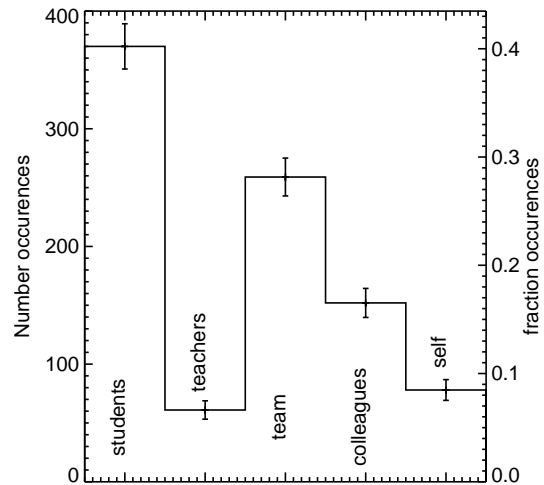}
\caption{Number of instances (left axis) and rate (right axis) of each of the five encoding words used to probe inward/outward facing stance of educators. {\em Students}-370 (40\%), {\em teachers}-61 (6.6\%), {\em team}-259 (28\%), {\em colleagues}-152 (17\%), {\em self}-78 (8.5\%). Uncertainties as shown correspond to the left axis and assume Poisson counting statistics; fractional uncertainties are $\leq$3\%, again assuming Poisson errors. {\em Students} and {\em teachers} are more outward-focused, and {\em colleagues} and {\em self} are more inward-focused. {\em Students} is, by far, the most common word in the encoding.
\label{fig:wordrate}}
\end{figure}

For each survey, the number of times these words appeared {\bf in the encoding} (not in the raw text from the surveys) were counted. The answers prior to encoding ranged from pithy to verbose, and the earlier years have fewer answers that could be mined in this fashion. For example, the recent surveys have 11 multi-part questions, but an educator from 2010 might only have provided partial answers to one question. To account for this diversity in answers, the fractional rate at which these encoded words occurred within the encoded surveys, over each person, team, year, or overall were calculated.  Over all the surveys, the rates shown in Fig.~\ref{fig:wordrate} were found; {\em students} is by far the most common word, and {\em team} is the second most common word. It is perhaps not surprising that {\em students} is so common, given that many educators enter the profession because of a desire to make a difference in students' lives. 

In order to quantify this continuum, a metric was developed here and calculated for each teacher. The number of times the encoding words were {\em colleagues} or {\em self} were divided by the number of total encoded words; subtracted from that was the number of instances of {\em students} or {\em teachers}, divided by the number of total words:
\begin{eqnarray}
s =  \frac{N_{\rm colleagues} + N_{\rm self}}{N_{\rm total words}}   - \frac{N_{\rm students} + N_{\rm teachers}}{N_{\rm total words} }
\end{eqnarray}
The value of $s$ varies between $-$1 and 1. Values of $s<$0 suggest a more outward focus, and $s >$0 suggests a more inward focus. The rate at which {\em team} appears in the encoding is also important, but not included in the calculation of $s$; see below.  


\subsection{Results}
\label{sec:spectrumresults}

\begin{figure}
\includegraphics[width=3in]{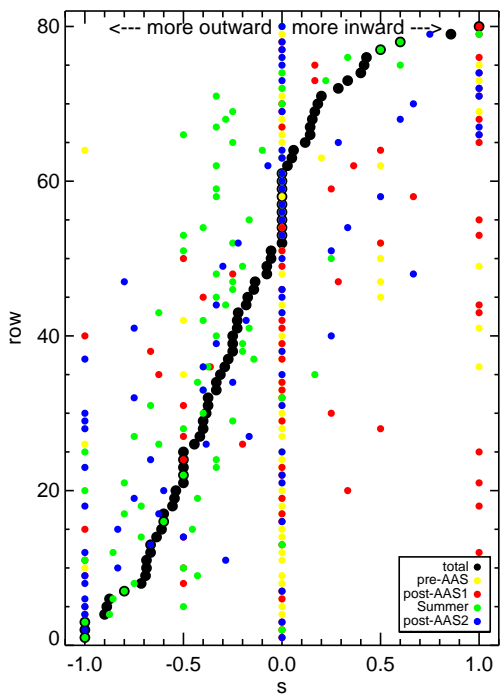}
\caption{Plot of metric, $s$ (see text). The $y$-axis is, in essence, an assigned teacher number; the teachers have been sorted by $s$ before plotting. Black points are the metric calculated over all instances of that educator for that team ($s_{\rm total}$); yellow points are the metric calculated from the pre-AAS survey (only available for the 2015 class and later), red points are the post-first-AAS survey (post-AAS1), green points are the summer survey, and blue points are the post-second-AAS survey (post-AAS2). The black points are larger than the colored points to make it clearer when the points are overlaid. Points more to the left reflect a more outward focus in the answers; points more to the right reflect a more inward focus. Every person has at least one survey to mine, so when $s_{\rm total}$ is exactly 0, it means that the existing (but typically limited) information suggests a balance between inward and outward motivations. However, many surveys are missing or insufficient from the first 4 years, so when $s$ at individual waypoints is 0, it is more often an indication that there is missing information. 
There are more red points on the far right and more blue points on the far left; see text. 
\label{fig:plotspectrum2}}
\end{figure}

Figure~\ref{fig:plotspectrum2} shows the distribution of the metric, $s_{\rm total}$, calculated over all surveys, for all the available data for each participant. To enhance understanding of the $s$ continuum, the educators are sorted by $s_{\rm total}$ before plotting. One row (one $y$-axis value) corresponds to one teacher. There are more than 74 rows because mentor educators may appear once on their first year of participation, and then again for each year of subsequent participation. Everyone has at least one survey to mine, so When $s_{\rm total}$ is exactly 0, then it is a real representation of the information that was in the existing surveys.

NITARP educators fall over the whole continuum range, and successful (as defined above) experiences are had by educators who fall over the whole range. The distribution of $s_{\rm total}$ over all the educators is slightly more biased towards the left ($<$0); more than twice as many educators are $<$0 (51) as are $>$0 (19). There are very few people who are mapped to exactly $+$1 or exactly $-1$ (1 and 3, respectively).  However, it is suspected that some people who were not selected for the program would likely end up as strongly one of these extremes. Experience collectors (\S\ref{sec:whyparticipate}) would likely emerge as $s=+1$ because they are focused on themselves and their comments would reflect that. Teachers who either are very frustrated and can't learn or choose not to learn would have $s=-1$ because one response to being faced with substantial learning challenges is to instead focus on others; educators who are overwhelmed and `shut down' may in some cases retreat to the familiar and focus on helping their students, which would result in $s=-1$.  These bins only serve to classify participants' experiences at these way points; these are descriptions of teachers' motivations, whether inward or outwardly focused and are not judgments.

\begin{quote}
{\em As a teacher who loves doing projects with students, I was in a much different role this time.  I have been taught to give quick help, activate students then move away as they engage.  When I often want to complete the task and do it for them, that wasn’t what the students needed for growth.  This time, I needed to stay engaged in the activity.  This might seem subtle, but it was not for me.  And usually, my personal projects are self-contrived.} -- NITARP educator, 2017 class 
\end{quote}

\begin{figure}
\includegraphics[width=3in]{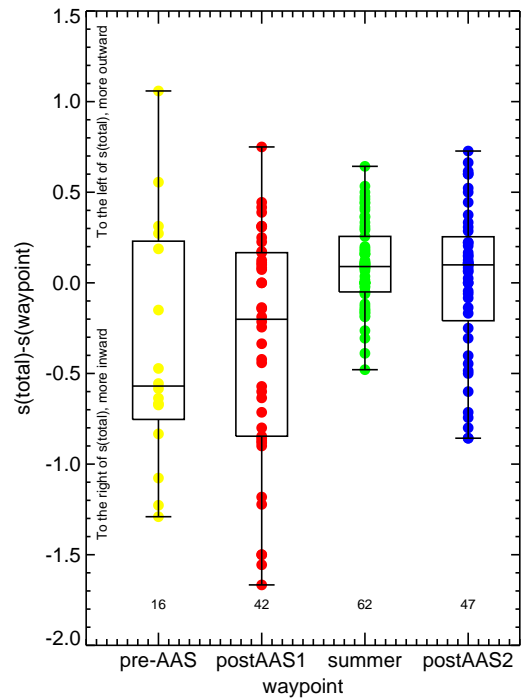}
\caption{Distribution of the offset in $s$ over the year; $s$ was calculated for each teacher at each waypoint ($s_{\rm waypoint}$) and subtracted from $s$ calculated over all available data for each teacher ($s_{\rm total}$). Values $s_{\rm total}-s_{\rm waypoint}<$0 indicate points to the right of $s_{\rm total}$ (e.g., more inward-focused at that waypoint than overall), and $s_{\rm total}-s_{\rm waypoint}>$0 are points to the left of $s_{\rm total}$ (e.g., more outward-focused at that waypoint than overall). Instances where $s_{\rm waypoint}$=0, where there is insufficient information to calculate $s$, have been omitted from this analysis. Total numbers of encoded surveys available at each waypoint are noted below each box (16, 42, 62, and 47 for the four waypoints, respectively). Box plots are shown on top of each distribution. The median values move up (teachers become more outward focused) with time during the NITARP year. 
\label{fig:plotspectrum3}}
\end{figure}

Figure~\ref{fig:plotspectrum2} also shows the time dependence of this metric; $s_{\rm waypoint}$ can be calculated separately for each waypoint.  In Fig.~\ref{fig:plotspectrum2}, red and yellow points are early in the year, green is midway through the year, and blue points are at the end of the year.  There are surveys missing (or missing sufficient information) for individual waypoints for some people, all from 2010-2013, because there is overall less information from those teachers. In those cases, $s_{\rm waypoint}$ is exactly 0.  

To better understand the evolution over time, $s$ was calculated for each teacher at each waypoint ($s_{\rm waypoint}$) and subtracted from $s$ calculated over all available data for each teacher ($s_{\rm total}$). (Instances where there is insufficient information to calculate $s_{\rm waypoint}$ have explicitly been omitted from this part of the analysis.) Values $<$0 indicate points to the right of $s_{\rm total}$ (e.g., more inward-focused at that waypoint than overall), and $>$0 are points to the left of $s_{\rm total}$ (e.g., more outward-focused at that waypoint than overall). The distributions of these changes in $s$ are shown in Figure~\ref{fig:plotspectrum3}. 

The offsets of the $s$ metric with time reflect at least in part how the program is structured. At the time of the pre-AAS surveys, they have just been selected for the program. In the application material, and during the online interviews, it is emphasized how this program is for their benefit, as teachers. The distribution of $s_{\rm pre-AAS}$ reflects that influence; the median offset in $s_{\rm pre-AAS}$ is the most inward-focused of all the waypoints. As part of the first AAS, the bootcamp emphasizes that the experience is primarily for them and secondarily for their students. But, during the rest of that first AAS, they have also met all the previous years' teachers and students who are finishing up, and are getting excited about the experience, including how they can share it with their students. Thus, the median offset in $s_{\rm post-AAS1}$ is more outward-focused than the median offset in $s_{\rm pre-AAS}$.  The summer visit is when the teams are working as teachers, students, and scientists, all side-by-side, and many people note this as a very positive thing; they particularly enjoy working towards a common goal in a community of equals made up of people from across the country.  Although the `team' encoding is not included in the $s$ metric, the words that the teachers use in their feedback forms reflect their thinking more about sharing with students and other (non-NITARP) teachers. Again, the median offset in $s_{\rm summer}$ is more outward-focused than the median offset in $s_{\rm post-AAS1}$ Finally, after the last AAS, the educators, with their students and the rest of their team, have stood by their poster and defended their research to other astronomers at the AAS; the median offset in $s_{\rm post-AAS2}$ is comparably outward-focused as the median offset in $s_{\rm summer}$. Most educators very much enjoy seeing the students increase in confidence while presenting their work. Their forms are frequently filled with references to their students, and sometimes moreover how this experience will affect their future students.  Note that for the educators who opted not to bring students to the second AAS, their feedback forms often still highlight the accomplishments of students brought by other teachers on the team.

\begin{quote}
{\em The greatest highlight of the week was listening to and watching my students collaborate with students from [the rest of the team...] What surprised me the most was the great sense of pride I felt when I listened to my students work with the other students in the [...] team. They completed each other's sentences. They interceded when others faltered. Wow.} -- NITARP educator, 2011 class
\end{quote}

\begin{quote}
{\em The most interesting part of my experience was how well our student teams bonded to successfully work, and play, together.  It was amazing and an important display of cooperative learning.  They did not hesitate to help each other as well as the teachers.} -- NITARP educator, 2014 class
\end{quote}

\begin{quote}
{\em The best thing about the trip was getting to spend time together working. I enjoyed watching the kids get to know each other and the other adults. I watched them grow in confidence and it made me feel very accomplished. } -- NITARP educator, 2016 class
\end{quote}

\begin{quote}
{\em I really enjoyed watching the ways that the older members of our team supported and interacted with the younger students.  They really bonded together better than I expected given the range of ages and skills.  I was also impressed by the number of people that came to talk with the students and really engaged with them in conversations about the science and process related to their poster and talked with them about their experience.  Astronomy is a wonderfully supportive community. } -- NITARP educator, 2016 class
\end{quote}

It is also likely to be the case that by the end of the program, the educators are more confident in their knowledge -- content and approach -- and in their ability to convey the information to their students. This would also result in more frequent `student' references at the post-second-AAS waypoint.

Most of the participating teachers move around near the middle of the distribution. However, note that some teachers can change substantially over the year (Fig.~\ref{fig:plotspectrum2}). Note, too, that those teachers who have extreme $s_{\rm total}$ values (near $+1$ or $-1$) tend to have comparable $s_{\rm waypoint}$ values throughout the year. That is, teachers who have $s_{\rm total}$ near $-1$ tend to have no  $s_{\rm waypoint}$ values near $+1$ (and the reverse is also true; those with  $s_{\rm total}$ near $+1$ tend to have no $s_{\rm waypoint}$ values near $-1$). Additionally, the majority of the most extreme $s_{\rm total}$ values are those from the earliest years of the program. The lack of extreme $s_{\rm total}$ values in more recent years probably reflects both that better questions are being asked in the surveys and that the program has the luxury of selecting from a rich applicant pool, so those teachers who are likely to have very extreme $s_{\rm total}$ values are now avoided.

\begin{figure}
\includegraphics[width=3in]{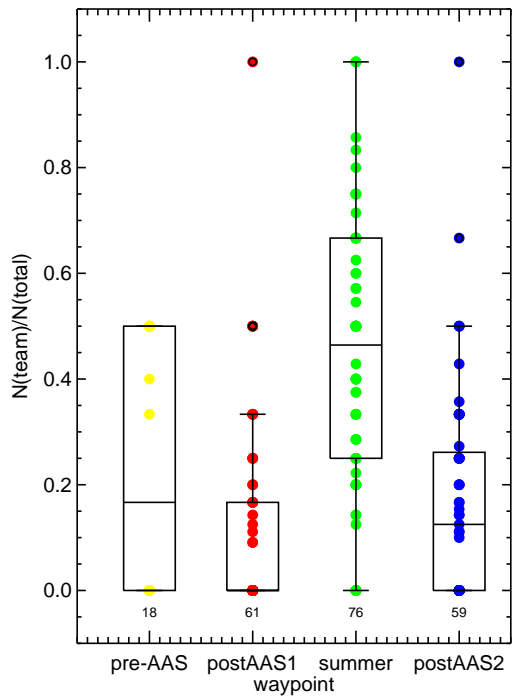}
\caption{Plot of the distribution of the fraction of {\em team} encoded words at each of the four waypoints. Notation is as in Fig.~\ref{fig:plotspectrum3}. Significantly more people focus on team-related comments after the summer visit than at any other time.
\label{fig:teamspectrum3}}
\end{figure}

To this point in this section, there is no inclusion of the influence of the word {\em team} in the encoded responses.  There is a strong time dependence of the {\em team} frequency; see Fig.~\ref{fig:teamspectrum3}. The summer meeting is when the entire group comes together for an intensive work week. It is not surprising, therefore, that a much larger fraction of their comments at that point focus on their team at the summer waypoint than at any other time.

\begin{quote}
{\em [The best thing was] Working with everyone. It was a great experience to work with the whole team from around the country in conjunction with the astronomer. The team working, collaboration and excitement was contagious and motivating. We had such an amazing experience working together. This has truly been on of the most amazing trips simply because we had an amazing group of students, teachers and astronomer to work with. } -- NITARP educator, 2013 class 
\end{quote}

\begin{quote}
{\em It was important to work on the data together, as a large team -- we were able to pair with new partners and better understand the questions we had. } -- NITARP educator, 2015 class
\end{quote}

\begin{quote}
{\em Getting to be part of a team of astronomers doing science has always been a dream of mine.} -- NITARP educator, 2017 class
\end{quote}

\begin{quote}
{\em I also benefited greatly from the combination of educators in the room. When something was introduced that I did not get the first time...one of the other educators could help. We all had different things we did well and could use set of resources to support each other and the students. } -- NITARP educator, 2013 class
\end{quote}

\begin{quote}
{\em I keep coming back to the team building.  To truly make this a successful collaborative effort between all of the teachers and all of the students and to get the commitment to the project it is vital that everyone meet in the same place and work together on site. } -- NITARP educator, 2015 class 
\end{quote}

\subsection{Implications}
\label{sec:spectrumimplications}

Authentic-research-based PD programs for science teachers can be very successful for teachers who fall anywhere over the whole range of this inward/outward focus continuum. NITARP educators similarly fall over the whole range of this continuum. The program is most geared towards helping teachers who want to learn themselves, either alongside their students or before sharing the information with their students, so there has most likely been an implicit emphasis on inward-focused educators. The program does not specifically select teachers based upon position along along the continuum; however, in retrospect, the program does not generally select participants that would emerge as strongly either of the extremes. For example, some teachers apply to NITARP explicitly stating that they want to send students to ISEF or want help starting a research class at their school. The program is not really able to help teachers with individual students' ISEF projects; the projects undertaken by NITARP teams are thoroughly group efforts (and therefore ineligible for ISEF). The program is also not really able to help individual students/teachers with developing academic-year-long research classes at their schools; NITARP spans a calendar year, not an academic year (and therefore inappropriate for merging directly to a school-year-long research class), and moreover, professional astronomers are most qualified and able to help with astronomy research, not class development.  NITARP's focus is teacher learning, which is emphasized in the application/interview process; given the relatively new introduction of the interview itself into the application process, there may have been a new bias introduced against those outward-focused teachers who are not so extremely outward-focused that they couldn't benefit from the experience.  

This program is both highly intensive and selective, and the participants are generally highly motivated. Nonetheless, because the participants are thrown into situations where they are learning difficult concepts, they cannot be expected to just `figure it out' all on their own; they must be supported as they are learning. The analysis in this paper has implications for how the program can best support participants through their experience.  Outward-focused teachers might be supported over frustrating junctures by relating to them at the level of how this benefits their current and future students.  Those educators that are strongly inward focused may be particularly susceptible to the `big fish/small pond' problem (Sec.~\ref{sec:fish}). If they are participating primarily because they see themselves as `big fish,' then the realization that they are not keeping up (and may not [yet] know how to keep up) may have a significant negative impact. They may not be able to salvage enough motivation from student-related gains to maintain their own participation in the program. The `big fish/small pond' problem is already addressed head-on (as discussed above); evidently, in supporting those participants, there may be a need to place less emphasis on student gains.

In order to meet the needs of both inward- and outward-focused educators, even as the program and educators' needs shift over time, mentor scientists and mentor educators have to be aware of these issues, even if the $s$ metric is not robust. Even if the specific location on the continuum where any given teacher starts the year, and how they move, is not well-parameterized by $s$, being aware of these issues can provide insight as to when during the year the teacher needs additional support, and how it might be provided. Educators across the continuum find that the learning experience provided is one of the primary rewards for participating in the program. However, if an educator is more outwardly focused, periodic reminders to pay attention to their own learning may be warranted; if an educator is more inwardly focused, it may be worth providing explicit prompts to reflect on how the experience may impact their teaching practices.

As discussed in Section~\ref{sec:successfulpd}, NITARP shares many qualities of successful PD. If other PD providers wish to create a PD experience like NITARP, following the NITARP structure, then PD providers also need to be aware of the continuum discussed here, and how best to support educators through the experience. For other PD programs that provide sustained interaction and a significant amount of work required to accomplish the goals of the program, it is also possible that participants' motivations to participate will change over the duration of the program. This is not something that was particularly anticipated; it would be easy to assume that someone undertaking PD requiring significant effort participates for primarily one or a few reasons that persist through their participation in the program. However, this is not what we observed here.

\subsection{Limitations}

This work is based on self-reported data from teacher participants. It is triangulated between multiple surveys from the same person at different waypoints, as well as multiple surveys at the same waypoint from different people. However, the surveys were not designed to place the participants on this continuum (since the continuum presented here was only recognized in the context of this work); the participants have no stake in answering the questions in a particular way to place themselves on the continuum. 

Accumulating more surveys over more years will increase the sample size. However, NITARP is changing with time in response to what is learned from these surveys and from suggestions made by participants (teachers and students) during and after each year. Working with similar themes in student data obtained concurrently with the teachers is beyond the scope of this work.

The $s$ metric is defined based on word counting of encoded words from the encoded responses. This is two steps away from the words written by the educators, but it compensates for significant length differences in responses (6-word partial sentence vs.\ 250+ word multiple paragraphs in response to the same question). It allows us to capture the overall tone of the response, as opposed to depending on respondents having used one of the five encoded words we used here to represent the emergent themes in the answers.

This $s$ metric may be affected by the context in which the teachers are filling out the surveys and is based on a limited number of survey responses, particularly in the first four years analyzed here. However, there is still insight, albeit potentially tentative, to be found about the participants and their motivations. 

Other programs may benefit from applying the metric discussed here, or a similar kind of metric. Insight into how best to support educators during their work in an intensive program like NITARP can help all participants feel like they are successful during their experience, as well as actually be successful. It is important to note, however, that this program is both highly intensive and selective, and such a metric might provide different results for teachers who are not as highly motivated as NITARP participants. 

The ramifications of an inward/outward focused teacher on the classroom is beyond the scope of this work.


\section{Summary and Future Work}
\label{sec:summary}

As ongoing change \cite{standards2013next,naframework} to science education in the US continues, there is increasing demand for high-quality PD that includes authentic research experiences for teachers. The NGSS calls for students to engage in the practice of science. It is difficult for teachers to engage students at this level without having experienced it themselves. Many educators have not yet had the opportunity to engage in authentic science research before being in the classroom; such experiences have the potential to be transformative for the educators \cite{herrington2016want}. 

NITARP, the NASA/IPAC Teacher Archive Research Program, has been partnering small groups of mostly high school classroom teachers with a research astronomer for a year-long authentic astronomy research project since 2005, working with a total of 103 educators from 34 states. The empirical data used in the qualitative and quantitative analysis here focuses on the last eight years (2010-2017) of surveys collected at up to 4 waypoints from 74 educator participants. 

NITARP aligns with many literature-identified characteristics of successful PD, including sustained interaction, creation of a community of practice and ongoing support of participants, a high level of rigor, and participants being treated as professionals. 

The original research question of this work was: How do participating teachers describe their motivations for participating in NITARP as evidenced in their feedback forms?  

Teachers participate for a variety of reasons, which were assessed from the word choice they used on their surveys. A metric was developed which allows participants to be mapped onto an inward-focused/outward-focused continuum. Inward-focused educators tend to have more personal learning goals, and outward-focused educators tend to have more teaching goals for their students or fellow teachers; both sides of the continuum are valued. Successful participants can be identified over the whole range of this continuum, and moreover they move during the year.  Identification of a teacher's focus and how it changes over time has implications for how PD programs such as NITARP can best support their participants, especially through junctures of frustration (see also Burrows \etal~\cite{burrows2016authentic}). NITARP was originally structured to more easily work with those with an inward focus; the results described here provide insight into how to help all participants, but perhaps more critically those with an outward focus, over the difficult parts of their experience.

There has not been very much systematic work on teacher research experiences \cite{buxner2014exploring,sadler2010}. This work adds to the existing body of literature to showcase snapshots of teachers' motivation throughout their year-long NITARP experience. 

There are many opportunities for future research into the motivations of educators in participating in this program, such as the motivations of mentor educators (and mentor applicants) for participating on teams for multiple years, the motivations of alumni who raise their own money to continue to attend AAS meetings, and the long-term impact of NITARP on participants and their future PD opportunities (both those that they offer and those they attend).

\begin{acknowledgments}
Thank you to all 103 NITARP \& Spitzer educators for your tireless devotion to this program! 

Support for this program was provided in part by NASA/ADAP funds. Thank you Doug Hudgins!

Thanks to Martha Kirouac and Tim Spuck for useful suggestions on early paper drafts.
\end{acknowledgments}

\appendix

\section{Details of Encoding}
\label{app:encodingdetails}

This Appendix contains a more detailed explanation (with examples) of encoding.

The survey questions (App.~\ref{app:questions}) did not ask about inward or outward focus, or any other synonyms referencing this idea presented in this paper.  At no point was there a question resembling, ``What is your motivation for being here?'' This was a theme that emerged from reading all the surveys together, more than once. This theme emerged from the tone and focus of the answers on the feedback forms. 

Whatever answers existed for each person, for each survey, the essence of the teachers’ responses to these survey questions was encoded using the five words (students, teachers, team, colleagues, self). The definition of these encoded words is included in Sec.~\ref{sec:placementonthecontinuum}.  

For example, if an educator responded to the question, ``What was the most interesting thing you did/saw/learned?'' with words about how wonderful it was to watch his/her students gain in confidence over the program, then that would be encoded as {\em students}. These encoded words did not reflect the specific use of the word in the answers, but the tone of the response. The word {\em colleague} need not be in the prompt or even the answer for the answer to be encoded as {\em colleague.}  If the teacher said, in response to, ``What was the most surprising thing you learned?'' with ``I met so many people doing what I'm trying to do and I can learn from their experiences,'' then this was encoded as {\em colleague}, because they are talking about personal gains from interactions with colleagues. If the teacher said (in response to any question), ``I am looking forward to sharing with my colleagues when I get home,'' then it was NOT encoded as ``colleague''; it was encoded as {\em teachers}, because they are talking about sharing with other teachers.   As another example, in response to ``What was the most interesting thing you learned?'', many educators responded with words about how their students are pretty amazing and motivated people; this gets encoded as {\em students} despite the question being about what they learned. Thus, we take the words/tone/content used in surveys, and encode them into any combination of the 5 words. 

After the encoding, we took a word count {\em of the encoding}.  Someone who wrote 6 words in response to one question and talked only about their students would be encoded as one instance of {\em students.}  Someone who wrote 250 words in response to one question and still talked only about their students would also be encoded as just one instance of {\em students,} because it was in response to one question.  Someone who wrote 250 words in response to one question and talked about their own gains and watching their students learn would be encoded as {\em self; students.}  In practice, someone who wrote 250 words is more likely to provide enough information so as to be encoded with more than one of the five words for that question, but not always. 

There is up to one of each encoded word per question answered. For example, someone who wrote 500 words in response to one question, talking only about themselves, would be encoded as just one instance of {\em self}; someone who wrote 5 words in response to each of two questions, talking only about themselves, would be encoded as one instance of {\em self} for each question. Someone who replied to three of 10 questions could only be encoded for three questions.  In calculating the metric, we divide by the total number of encoded words (total for that person or for that person's survey, depending on what is being calculated) in order to at least partially compensate for missing answers or overly terse responses.

Because we know these people, we endeavored to not let our opinions of the individuals color our encoding, relying entirely on what they wrote. In several cases, their survey answers revealed a different focus than we might have assumed they had based just on memory.  For example, one person emerged as solidly `self' in the encoding, from all the surveys collected from that individual. Thinking back on all interactions with this person, this makes sense in retrospect, but we would not necessarily have put this person in this bin {\em a priori}.

\section{Questions in the Surveys}
\label{app:questions}

\subsection{Prior to the First AAS}

We ask these questions after they are accepted into the program, prior to their coming to their first AAS. We started this with the 2015 class.

\begin{itemize}
\item What do you expect to gain from your NITARP experience?
\item What do you expect to learn at your first AAS meeting? 
\item What is ``real astronomy''?
\item What qualities do you think are important to be an astronomer?
\item How will you engage with other teachers on your team? 
\item What are your professional goals and career plans?
\end{itemize}

\subsection{After the AAS}

In 2010, we started by creating a worksheet that was designed to help the educators make sense of the chaos that is the AAS. We gave them specific tasks covering all of the major reasons why professional astronomers go to the AAS. This worked in that it gave them explicit tasks to accomplish at the AAS, but it also meant that most participants focused on those tasks, and didn't give us detailed answers that would give us insight into, e.g., their reasons for participation in NITARP.

The tasks in this original worksheet covered the following (with many more details given in the worksheet than are listed here):
\begin{itemize}
\item Networking -- find educators not in NITARP
\item Your science --  attend talks about your research topic 
\item New science -- attend talks about other kinds of science, not your topic
\item Policy -- attend a town hall 
\item Observatories -- find booths run by observatories 
\item Industry -- find booths run by industry
\item Publishers -- find booths run by publishers
\item Education -- attend an education session
\item Posters -- find all the NITARP-affiliated posters at the meeting
\end{itemize}

The worksheet from after the 2010, 2011, and 2012 AAS meetings stopped here. The worksheet from after the 2013 and 2014 AAS meetings continued with (the `Sunday workshop' is the NITARP Bootcamp):
\begin{quote} 
Finally...Putting it all together: (FOLKS FINISHING UP: consider the entire NITARP experience when answering these; NEW FOLKS: consider just the AAS+Sunday workshop when answering these.) What was the most interesting thing you did/saw/learned? Was there anything that happened that you did not anticipate? Did this experience change the way you thought about astronomy or astronomers or NITARP?  How is this going to change the way you work in the classroom (as an educator or student)? Is there any advice you'd give the folks who are coming next time? Or, advice you'd give us as the people running NITARP?
\end{quote}
Because this was at the end of the worksheet, we found that many participants skipped these questions, or only answered some of them. 

As part of the mid-2014 overhaul of all our surveys, we separated the AAS worksheet itself (which several participants had explicitly said they valued as a way to give shape to their AAS experience) from the feedback form. Since the 2015 AAS, then, the survey has looked like the below. Both the teams finishing and the teams starting are asked to answer the same questions; teachers and students answer the same questions. There is a preamble that asks the class finishing up to consider the entire NITARP experience when answering these, and the new class to consider just the AAS+NITARP Bootcamp -- their NITARP experiences so far -- when answering these.
\begin{itemize}
\item What was the most interesting thing you did/saw/learned? How did the reality of this experience compare to your expectations for what you would learn/do/see? 
\item Was there anything that happened that you did not anticipate? 
\item Did this experience (so far) change the way you thought about astronomy or astronomers or NITARP?  
\item What new resources did you learn about? How did the scientists teach you to use them?
\item Can you tell us about a time you got confused or frustrated during this experience? Did you work through it? Did you get enough support? What supports helped? Could you still use support? 
\item How is this going to change the way you work in the classroom? (For educators:  What have you changed or added to your teaching as a result of NITARP?)
\item What are your professional goals and career plans? Has this experience changed them?
\item How has this program impacted your thoughts on/plans for your next educational or professional development experience?
\item If you had to tell Congress what teachers or students learn/experience/start as a result of experiences such as NITARP, what would you say?
\item Is there any advice you'd give the folks who are coming next time? What advice would you pass on to the NITARP \item Could you have done this (entire NITARP project) alone in a reasonable amount of time? Why did guidance help? What kept you going during the research process?
\end{itemize}

\subsection{After the Summer Visit}

After the summer visit, we asked the 2010 class to answer these questions:
\begin{itemize}
\item What was the most important thing (or few things) you learned?
\item What was the most surprising thing you learned?
\item What was the least surprising thing you learned?
\item We know the travel arrangements were a nightmare, and we're trying to make sure that it goes more smoothly for the AAS and for next year's teams.  BUT, beyond that, is there anything that you would have had us do differently (or that you yourself would have done differently), knowing what you do now?
\item What was the best thing about the trip?
\item What did you do with the data associated with your project while you were here?
\item What do you plan to do with the data when you return home?
\item What is ``real astronomy''?  Did you do anything on this visit (or as part of this experience so far) that you expected would be part of scientific research? Or anything that you did not think would be part of scientific research?  Why or why not?
\end{itemize}

After the 2011 class, we asked very similar questions:
\begin{itemize}
\item What was the most important or interesting thing (or few things) you did/saw/learned?
\item What was the most surprising thing you did/saw/learned? Did anything happen that you did not anticipate?
\item What was the least surprising thing you did/saw/learned?
\item What was the best thing about the trip?
\item What, in broad terms, did you do with the data associated with your project while you were here?
\item What, in broad terms, do you plan to do with the data when you return home?
\item Did this experience (so far) change the way you thought about astronomy or astronomers?
\item What is ``real astronomy''?  Did you do anything on this visit that you expected would be part of scientific research? Or anything that you did not think would be part of scientific research?  Why or why not?
\item Do you have any advice for the teachers (or students) coming on visits after yours?
\item Regarding the travel arrangements -- is there anything that you would have had us do differently (or that you yourself would have done differently), knowing what you do now?
\end{itemize}

In summer 2012, we experimented with bringing in staff from the rest of IPAC to talk about their career path. It was not entirely successful. We added a single question to see if the teachers agreed: ``Was the `career lunch' where you met other people from across IPAC, Caltech, and/or JPL useful, e.g., should we do it again?''

The survey from summer 2013 and 2014 returned to that from the 2011 class.

Starting in Summer 2015, and running through 2017, these are the questions we asked:
\begin{itemize}
\item What was the most important or interesting thing (or few things) you did/saw/learned?
\item What was the most surprising thing you did/saw/learned? Did anything happen that you did not anticipate?
\item What was the least surprising thing you did/saw/learned?
\item What was the best thing about the trip?
\item Could the work you carried out during the visit be done online?  Were there any benefits of working together at Caltech? Did the group change after the visit? If so, how?
\item Can you tell us about a time you got confused or frustrated during the process? Did you work through it? Did you get enough support? What supports helped? Could you still use support? 
\item Did this experience (so far) change the way you thought about astronomy or astronomers?
\item What is ``real astronomy''?  Did you do anything on this visit that you expected would be part of scientific research? Or anything that you did not think would be part of scientific research?  Why or why not?
\item What qualities do you think are important to be an astronomer?
\item Do you have any advice for the teachers (or students) coming on visits after yours? Is there anything we could have done to improve your visit?
\end{itemize}


\bibliography{References.bib}

\end{document}